\documentclass[reprint, aps, superscriptaddress, twocolumn]{revtex4-2}

\usepackage{amsmath,amssymb,amsfonts,mathtools,bm}
\usepackage{graphicx}
\usepackage{booktabs}
\usepackage{orcidlink} 
\usepackage{xcolor}

\usepackage{hyperref}
\hypersetup{
  colorlinks=true,
  linkcolor=blue,
  citecolor=blue,
  urlcolor=blue,
}

\begin{document}

\title{A phenomenological decomposition of SRC-inspired occupation effects in imbalanced Fermi superfluids}

\author{Lucas A.~Souza\,\orcidlink{0000-0002-8711-956X}}
\email{lasouza@if.usp.br}
\affiliation{Minas Gerais, Brazil}

\begin{abstract}
Imbalanced two-component Fermi superfluids are usually formulated with
sharp Fermi-step occupations, although interacting systems exhibit depletion
below the Fermi surface and a high-momentum tail. We introduce a
sector-resolved phenomenological framework in which one prescribed
short-range-correlation (SRC)-inspired momentum distribution modifies either
the anomalous pairing contribution (model P), the imbalance-induced blocking
term (model B), or both (model P+B). The four realizations---ideal, P, B, and
P+B---are compared through the grand-potential landscape and through a
fixed-density reconstruction of the paired/phase-separated to polarized-normal
boundary in three and two dimensions. In both dimensionalities, the
pairing-sector modification accounts for nearly all of the reduction in the
critical polarization, whereas the blocking-sector correction remains
subdominant. Consequently, the phase boundaries form two nearly degenerate
pairs, ideal $\simeq$ B and P $\simeq$ P+B. These results quantify, within
the stated phenomenological construction, how a prescribed SRC-inspired
occupation propagates separately through the pairing and blocking sectors of
an imbalanced superfluid.
\end{abstract}

\maketitle

\section{Introduction}
\label{sec:introduction}

A two-component Fermi gas with unequal populations provides a direct
setting in which pairing, polarization, and phase separation compete. At
small chemical-potential or population imbalance, pairing favors a
superfluid state; as the mismatch between the two Fermi surfaces grows, the
system can instead become polarized and normal or separate into paired and
normal regions. This competition was established in three dimensions by
Bedaque, Caldas, and Rupak \cite{Bedaque2003}; the corresponding
zero-temperature superfluid--normal quantum transition and its critical
imbalance have also been discussed explicitly by Caldas \cite{Caldas2020}.
The problem has since been studied in
mean-field and beyond-mean-field approaches
\cite{Sheehy2006,Hu2007,He2008,Strack2014,Wu2016,Dutta2016,Tempere2008,Radzihovsky2010},
as well as experimentally \cite{Shin2006,Mitra2016}. Its two-dimensional
realization introduces additional structure associated with the 2D bound
state and the BCS--BEC evolution
\cite{Caldas2012,Melo2024}. In both dimensionalities, the phase boundary is
ultimately a fixed-density competition among paired, polarized normal, and
phase-separated configurations.

Most treatments of this problem adopt ideal step-function occupations,
$n_\sigma(k)=\Theta(k_{F\sigma}-k)$. Interacting Fermi systems, however,
show depletion below the Fermi momentum and nonzero occupation above it,
including high-momentum components associated with short-range
correlations. This structure is encoded in Tan's contact relations
\cite{TAN20082987,TAN20082971,TAN20082952}, observed in nuclear and
ultracold-atom systems \cite{Hen2015,Decamp2016,Daix2026}, and remains an
active subject in modern descriptions of universal and scale-separated
momentum distributions \cite{Tropiano2021,Rubtsova2024,Cai2026}.

Occupation-based phenomenological treatments incorporate correlation
effects by replacing the ideal Fermi step with a depleted distribution
containing high-momentum strength. Applied within a fixed dynamical
framework, this procedure isolates the response associated with the
momentum occupation itself. It has been used to study SRC-induced
changes in nuclear equations of state, neutron-star radii, tidal
deformabilities, and related observables
\cite{Souza2020,Rodrigues2023,Gautam2025}. Here we apply the same
comparative strategy to imbalanced pairing, evaluating a common
thermodynamic functional with ideal and SRC-inspired occupations.

The connection between correlated occupations and superfluid pairing is
particularly relevant here. Nuclear-matter calculations show that spectral
strength redistribution and high-momentum components can substantially
reduce pairing gaps relative to ideal-occupation results
\cite{Rios2017,Drissi2022,Sedrakian2019}. Recent work also emphasizes that
pairing and high-momentum structure are dynamically intertwined rather than
independent ingredients \cite{Li2026}. These results motivate asking how an
SRC-inspired occupation modifies not only the balanced pairing gap, but
also the competition between pairing and population imbalance.

In an imbalanced system, the same momentum distribution enters the problem
through two physically distinct sectors. First, depletion around the Fermi
surfaces changes the occupation-dependent weight of the anomalous pairing
contribution. Second, the occupation difference between the two species
changes the effective weight of the momentum interval blocked by the
imbalance. We represent these responses by a pairing-sector prescription
(P) and a blocking-sector prescription (B), both generated from the same
correlated occupation and both reducing exactly to the ideal theory in the
sharp-step limit. The construction is therefore a sector-resolved
phenomenological implementation of a common SRC-inspired input.

We compare four realizations of the same thermodynamic framework: the
ideal reference, the pairing-sector model P, the blocking-sector model
B, and the combined model P+B. This decomposition measures how the
change in critical polarization is distributed between the anomalous
and blocked contributions. The P correction acts over the anomalous
momentum domain, while B becomes active only when an imbalance-induced
blocked interval is present. The calculation determines the magnitude
and sign of each contribution, their combined effect, and the
persistence of the resulting hierarchy between two and three
dimensions.

The analysis is performed independently in three and two dimensions. The
three-dimensional formulation follows the regularization of
Ref.~\cite{Bedaque2003}, while the two-dimensional calculation follows the
contact-interaction treatment of Ref.~\cite{Caldas2012}. The latter is
placed in the broader modern context of two-dimensional BCS--BEC physics
reviewed in Ref.~\cite{Melo2024}. In both cases the grand-potential
landscape is supplemented by a full fixed-density reconstruction, so the
reported transition is the canonical boundary between the
paired/phase-separated state and the polarized normal state within the
model space considered.

The remainder of this paper is organized as follows. Section~\ref{sec:framework}
develops the ideal three- and two-dimensional references, the correlated
occupation, and the P/B sector decomposition. Section~\ref{sec:numerics}
summarizes the numerical implementation and convergence tests.
Section~\ref{sec:results} presents the grand-potential landscapes and the
canonical phase boundaries, and Sec.~\ref{sec:discussion} discusses how the
same SRC-inspired occupation propagates through the pairing and blocking
sectors in the two dimensionalities.

\section{Theoretical framework}
\label{sec:framework}

\subsection{Imbalanced two-component Fermi system}
\label{sec:system}

We consider a two-component Fermi gas with species $A$ and $B$,
equal mass $M$, and chemical potentials $\mu_A$ and $\mu_B$.
Throughout the calculation we set $\hbar=1$ and use $M=1$ in the
numerical implementation. The average chemical potential and the
imbalance field are defined as
\begin{equation}
\mu=\frac{\mu_A+\mu_B}{2},
\qquad
h=\frac{\mu_B-\mu_A}{2},
\label{eq:muh}
\end{equation}
Thus $\mu_A=\mu-h$ and $\mu_B=\mu+h$. The single-particle dispersion measured relative to the average chemical potential is
\begin{equation}
\xi_k=\frac{k^2}{2M}-\mu,
\label{eq:xik}
\end{equation}
common to both species once written in terms of $\mu$ and $h$. Pairing
between the two species is described by a single complex order parameter
$\Delta$, and the resulting Bogoliubov quasiparticle energy is
\begin{equation}
E_k=\sqrt{\xi_k^2+\Delta^2}.
\end{equation}
For homogeneous fixed-$(\mu,h)$ calculations, energies are expressed
relative to the average chemical potential $\mu$, momenta relative to
$\sqrt{2M\mu}$, and grand-potential densities in the corresponding
natural units. For the representative landscapes in Fig.~\ref{fig:omega-grid}, we use
$\mu=1$ in 3D. In 2D, $\mu=0.9841$ follows from the balanced mean-field
relations $\mu=\epsilon_F-|\epsilon_B|/2$ and
$\Delta_0^2=2\epsilon_F|\epsilon_B|$, with
$\Delta_0\simeq0.252$ and $\epsilon_F=1$.
These choices set only the plotting scale and do not enter the canonical
phase-boundary calculations.

In the equal-mass, fixed-$(\mu,h)$ representation the two quasiparticle
branches are $E_k-h$ and $E_k+h$. The lower branch, $E_k-h$, can become
negative on a finite momentum interval whenever
\begin{equation}
E_k<h
\quad\Longleftrightarrow\quad
|\xi_k|<\sqrt{h^2-\Delta^2},
\label{eq:blocked-condition}
\end{equation}
which requires $h>\Delta$; this is the \emph{blocked interval} responsible
for Pauli blocking of pairing in the imbalanced system.

The homogeneous branches are first obtained from the grand potential
$\Omega(\mu,h,\Delta)$. To determine the physical boundary at fixed
particle numbers, these branches are subsequently transformed to the
canonical ensemble and compared with the corresponding mixed
configuration. The distinction is essential: the stationary structure
of $\Omega$ at fixed chemical potentials identifies the available
homogeneous branches, whereas the critical polarization follows from
their fixed-density energy competition.

\subsection{Ideal mean-field reference}
\label{sec:ideal}

Before introducing correlated occupations we fix the ideal (step-occupation)
reference in each dimensionality; every model in Sec.~\ref{sec:src} reduces
to this limit when the occupation is replaced by a sharp Fermi step,
$n_\sigma(k)=\Theta(k_{F\sigma}-k)$.

\paragraph{Three dimensions.} Following Bedaque, Caldas, and Rupak
\cite{Bedaque2003}, the cutoff-regularized grand potential is
\begin{equation}
\Omega_{3D}(\Delta)
=
\frac{\Delta^2}{g_{\rm eff}}
+
\int_0^{\Lambda}\frac{k^2dk}{2\pi^2}
\Bigl[
\xi_k-E_k-(h-E_k)\,\Theta(h-E_k)
\Bigr].
\label{eq:omega3d-ideal}
\end{equation} 
Here $\Lambda$ is the ultraviolet momentum cutoff of the contact theory.
It is chosen well above the momentum scales associated with $\mu$, $h$,
and $\Delta$. After the bare coupling is adjusted consistently with
$\Lambda$, low-energy observables must become insensitive to further
increases of the cutoff.

Rather than varying the cutoff-dependent coupling directly, we use the
balanced reference gap $\Delta_0$ as the control parameter. For each
chosen $\Delta_0$, $g_{\rm eff}(\Lambda)$ is fixed by requiring that the
$h=0$ grand potential have a stationary minimum at $\Delta=\Delta_0$:
\begin{equation}
\frac{1}{g_{\rm eff}(\Lambda)}
=
\int_0^{\Lambda}\frac{k^2\,dk}{2\pi^2}\,
\frac{1}{2\sqrt{\xi_k^2+\Delta_0^2}}.
\label{eq:geff}
\end{equation}
Thus, $\Delta_0$ parametrizes the interaction scan performed in three
dimensions.

To express the same interaction in terms of the vacuum $s$-wave
scattering length $a$, the cutoff-dependent coupling must be related to
the renormalized contact interaction. For a sharp momentum cutoff,
\begin{equation}
\frac{1}{g_{\rm eff}(\Lambda)}
=
-\frac{M}{2\pi a}
+
\int_0^\Lambda \frac{k^2\,dk}{2\pi^2}\,
\frac{1}{2\epsilon_k},
\qquad
\epsilon_k=\frac{k^2}{2M},
\label{eq:geff-renorm}
\end{equation}
or, after evaluating the vacuum integral,
\begin{equation}
\frac{1}{g_{\rm eff}(\Lambda)}
=
-\frac{M}{2\pi a}
+
\frac{M\Lambda}{2\pi^2}.
\end{equation}
Combining this relation with Eq.~\eqref{eq:geff} gives
\begin{equation}
\frac{1}{k_Fa}
=
\frac{\Lambda}{\pi k_F}
-
\frac{2\pi}{M k_F\,g_{\rm eff}(\Lambda)}.
\label{eq:invkfa}
\end{equation}
Equation~\eqref{eq:invkfa} is used to convert the interaction values
defined by $\Delta_0$ into the standard scattering-length variable
$1/(k_Fa)$. The numerical implementation uses $M=1$. 

\paragraph{Two dimensions.} Following Caldas, Mota, Farias, and Souza
\cite{Caldas2012}, the two-body contact interaction is regularized directly
by the two-body binding energy $\epsilon_B$ rather than by a coupling
constant, giving the radial grand potential
\begin{widetext}
\begin{equation}
\Omega_{2D}(\Delta)
=
\int_0^{\Lambda}\frac{k\,dk}{2\pi}
\left[
\xi_k-E_k+\frac{\Delta^2}{2\epsilon_k+|\epsilon_B|}
-(h-E_k)\Theta(h-E_k)
\right],
\qquad
\epsilon_k\equiv\frac{k^2}{2M}.
\label{eq:omega2d-ideal}
\end{equation}
\end{widetext}
Here $\epsilon_k$ is the free (vacuum) dispersion, distinct from the
chemical-potential-shifted $\xi_k=\epsilon_k-\mu$ of
Eq.~\eqref{eq:xik}: the regularizing counterterm is fixed by the
vacuum two-body binding energy $\epsilon_B$ and is therefore
$\mu$-independent, matching the standard 2D contact-interaction
regularization. The interaction is already renormalized by the vacuum binding energy, so
$\Lambda$ is only a numerical truncation of the convergent integral. The
choice $\Lambda=16$ is conservative; Sec.~\ref{sec:numerics} verifies that
the reported observables are unchanged when the integration range is
extended.

\paragraph{Homogeneous branches.} In both dimensionalities, three
stationary structures of $\Omega(\Delta)$ are relevant: the BCS solution
(a minimum at $\Delta\neq0$), the normal solution ($\Delta=0$), and a
Sarma-like solution (an interior stationary point, generically a local
\emph{maximum} of $\Omega(\Delta)$ once $h>0$). Consistent with
Refs.~\cite{Bedaque2003,Caldas2012}, the Sarma branch is retained only as a
homogeneous-stability diagnostic; it is not treated as a candidate ground
state at fixed density, where the mixed state of Sec.~\ref{sec:mixed}
generally dominates.

\subsection{SRC-inspired correlated occupations}
\label{sec:src}

The correlated occupation provides the common input to the P and B
sectors. Its shape contains depletion below the Fermi momentum, a
reduced discontinuity near $k_F$, and an asymptotic $k^{-4}$ tail, as
motivated by microscopic calculations and contact relations
\cite{TAN20082987,TAN20082971,TAN20082952,Tropiano2021,Rubtsova2024,Cai2026}.
The ideal and SRC-inspired dimensionless profiles are defined below and
scaled by the species-dependent momenta $k_{F\sigma}$. During a given
calculation, the profile parameters are fixed while $k_{F\sigma}$ varies
with $(\mu,h)$.
The prescribed occupations enter the functional only through the pairing
weight $\Phi(k)$ and the blocking weight $B_{\rm SRC}(k)$ defined below.
This allows the same correlated profile to modify the pairing and blocking
contributions separately, while the interaction and the remaining
quasiparticle structure are kept unchanged. Thermodynamic densities are
obtained independently from derivatives of the grand potential. Because
the scales $k_{F\sigma}$ depend on the chemical potentials and the grand
potential also contains pairing coherence factors, the integrated
prescribed occupations need not coincide exactly with the corresponding
thermodynamic densities.

\subsubsection{Momentum-distribution model}
\label{sec:nmodel}

The correlated occupation is modeled as a piecewise-defined profile
with depletion below the Fermi momentum, a reduced discontinuity at
$q=1$, and a high-momentum $k^{-4}$ tail, 
consistent with Tan-contact scaling. For species $\sigma$ we write
$q=k/k_{F\sigma}$, where
\begin{equation}
k_{F\sigma}=\sqrt{2M\mu_\sigma},
\qquad
\mu_A=\mu-h,\quad \mu_B=\mu+h.
\label{eq:kfsigma-density}
\end{equation}
Because $k_{F\sigma}$ depends on $(\mu,h)$, the profiles
$n_\sigma(k)=n_{\rm profile}^{(d)}(k/k_{F\sigma})$
shift under the thermodynamic derivatives
$-\partial\Omega/\partial\mu$ and $-\partial\Omega/\partial h$.
 Their dimensionless form is held fixed
during the calculation.

For the ideal reference, the dimensionless occupation is the sharp Fermi
step,
\begin{equation}
n_{\rm ideal}(q)=\Theta(1-q).
\label{eq:nideal}
\end{equation}
For the SRC-inspired profiles, we use the piecewise form
\begin{equation}
n_{\rm corr}^{(d)}(q)=
\begin{cases}
\mathrm{clip}\!\left(
n_0-(n_0-n_{1-})q^\alpha,\,0,1
\right),
& q<1,\\[5pt]
\mathrm{clip}\!\left(
(n_{1+}-c_d)q^{-\beta}+c_d q^{-4},\,0,1
\right),
& q\geq1,
\end{cases}
\label{eq:nmodel}
\end{equation}
where $d=2,3$ labels the dimensionality and
$\mathrm{clip}(y,0,1)\equiv\min[1,\max(0,y)]$ restricts the occupation
to the interval $[0,1]$. 
For all parameter sets in Table~\ref{tab:benchmarks}, the unclipped
expressions already remain within $[0,1]$; the clipping operation is
therefore inactive and does not alter the normalization below.

The coefficient $c_d$ fixes the normalization of the correlated profile.
Defining
\begin{equation}
I_{\rm below}^{(d)}
=
\frac{n_0}{d}
-
\frac{n_0-n_{1-}}{\alpha+d},
\qquad
c_d
=
\frac{
\displaystyle \frac{1}{d}
-
I_{\rm below}^{(d)}
-
\frac{n_{1+}}{\beta-d}
}{
\displaystyle \frac{1}{4-d}
-
\frac{1}{\beta-d}
},
\label{eq:nmodel-constants}
\end{equation}
gives
\begin{equation}
\int_0^\infty n_{\rm corr}^{(d)}(q)\,q^{d-1}\,dq
=
\frac{1}{d},
\label{eq:nmodel-normalization}
\end{equation}
corresponding to the radial measures $k\,dk$ in 2D and $k^2dk$ in 3D.

For either the ideal or correlated choice, the per-species occupations are
\begin{equation}
n_\sigma(k)
=
n_{\rm profile}^{(d)}
\!\left(\frac{k}{k_{F\sigma}}\right),
\qquad
\sigma=A,B,
\label{eq:species-occupations}
\end{equation}
where $n_{\rm profile}^{(d)}=n_{\rm ideal}$ for the ideal model and
$n_{\rm profile}^{(d)}=n_{\rm corr}^{(d)}$ for the SRC-inspired models.

The parameters $n_0$, $n_{1-}$, and $n_{1+}$ determine the depletion
below $k_F$ and the reduced discontinuity at $q=1$, while $\alpha$ and
$\beta$ control the momentum dependence below and above the Fermi surface.
The explicit $c_dq^{-4}$ contribution fixes the asymptotic tail and
dominates for $q\gtrsim2$ in all nonideal profiles of
Table~\ref{tab:benchmarks}. The same values of
$(n_0,n_{1-},n_{1+},\alpha,\beta)$ are used in 2D and 3D; only $c_d$
changes to satisfy Eq.~\eqref{eq:nmodel-normalization}.

Table~\ref{tab:benchmarks} summarizes the profiles used in the
calculation. Their depletion and discontinuity ranges are guided by
correlated nuclear-matter and pairing studies
\cite{Rios2017,Drissi2022,Sedrakian2019} and are treated as a sensitivity
ensemble rather than as microscopic fits. The main results use the
\textit{NM-strong} profile, while the alternatives test the stability of
the sector hierarchy. The \textit{ideal} profile recovers the step
occupation of Sec.~\ref{sec:ideal}.

\begin{table}[h]
\centering
\caption{Reference correlated-occupation profiles entering
Eqs.~\eqref{eq:nmodel}--\eqref{eq:nmodel-normalization}. The
\textit{NM-strong} set is used for the main figures and quoted numerical
results, while \textit{NM-moderate} and \textit{SM-strong} are included
to test the sensitivity of the sector hierarchy. NM denotes
nuclear-matter-inspired depletion and SM a stronger schematic
modification; the ranges are guided by
Refs.~\cite{Rios2017,Drissi2022,Sedrakian2019} and are not intended as
microscopic fits.}
\label{tab:benchmarks}
\setlength{\tabcolsep}{7pt}
\renewcommand{\arraystretch}{1.08}
\begin{tabular}{lccccc}
\hline\hline
parameter set & $n_0$ & $n_{1-}$ & $n_{1+}$ & $\alpha$ & $\beta$ \\
\hline
ideal        & 1.00 & 1.00 & 0.00 & 2.0 & 8.0 \\
NM-moderate  & 0.97 & 0.90 & 0.05 & 2.0 & 8.0 \\
NM-strong    & 0.95 & 0.86 & 0.06 & 1.8 & 8.0 \\
SM-strong    & 0.88 & 0.74 & 0.10 & 1.2 & 8.0 \\
\hline\hline
\end{tabular}
\end{table}

\subsubsection{Pairing-coherence modification: Model P}
\label{sec:modelP}

The factor $1-2n_\sigma(k)$ measures the particle--hole occupation
imbalance available to the coherent pairing channel. We therefore use the
correlated occupations to define the symmetric, momentum-dependent measure
\begin{equation}
\Phi(k)=\sqrt{\,|1-2n_A(k)|\,|1-2n_B(k)|\,}\,,
\label{eq:Phi}
\end{equation}
which satisfies $\Phi(k)=1$ identically for the ideal step occupation. In
the grand-potential integrand, $\Phi(k)^2$ multiplies only the
\emph{anomalous}, gap-dependent contribution, not the normal-state part.
Writing $\xi_k^{0}\equiv|\xi_k|$, the pairing/normal sector of the
integrand is
\begin{equation}
\bigl(\xi_k-\xi_k^{0}\bigr)
+
\Phi(k)^2\bigl(\xi_k^{0}-E_k\bigr),
\label{eq:paired-sector}
\end{equation}
in place of the naive product $\Phi(k)(\xi_k-E_k)$. The two forms coincide
when $\Phi(k)=1$, recovering Eq.~\eqref{eq:omega3d-ideal} exactly. They
differ whenever $\Phi(k)<1$: Eq.~\eqref{eq:paired-sector} leaves the
$\Delta=0$ limit,
\begin{equation}
\left.\Bigl[(\xi_k-\xi_k^{0})+\Phi(k)^2(\xi_k^{0}-E_k)\Bigr]\right|_{\Delta=0}
=\xi_k-\xi_k^{0},
\end{equation}
exactly independent of $\Phi(k)$ (since $E_k\to\xi_k^{0}$ as
$\Delta\to0$), whereas the naive product does not. Model P is therefore deliberately defined as a sector-resolved probe:
the normal-state reference at $\Delta=0$ is left unchanged, while only the
anomalous condensation contribution is reweighted. The combination
$\Phi^2(k)=|1-2n_A(k)|\,|1-2n_B(k)|$ is symmetric under exchange of the two
species, reduces identically to unity for ideal occupations, and is
motivated by occupation-dependent coherence factors that enter correlated
pairing kernels, including the SRC treatment of Ref.~\cite{Rios2017}. The geometric mean treats both components on equal footing, remains
non-negative across the Fermi-surface discontinuities, and yields the product
$\Phi^2$ that weights the two-species anomalous channel. This construction
provides the operational pairing-sector realization of the prescribed
occupation used throughout the present framework.

\begin{figure}[!t]
    \centering
    \includegraphics[width=\columnwidth]{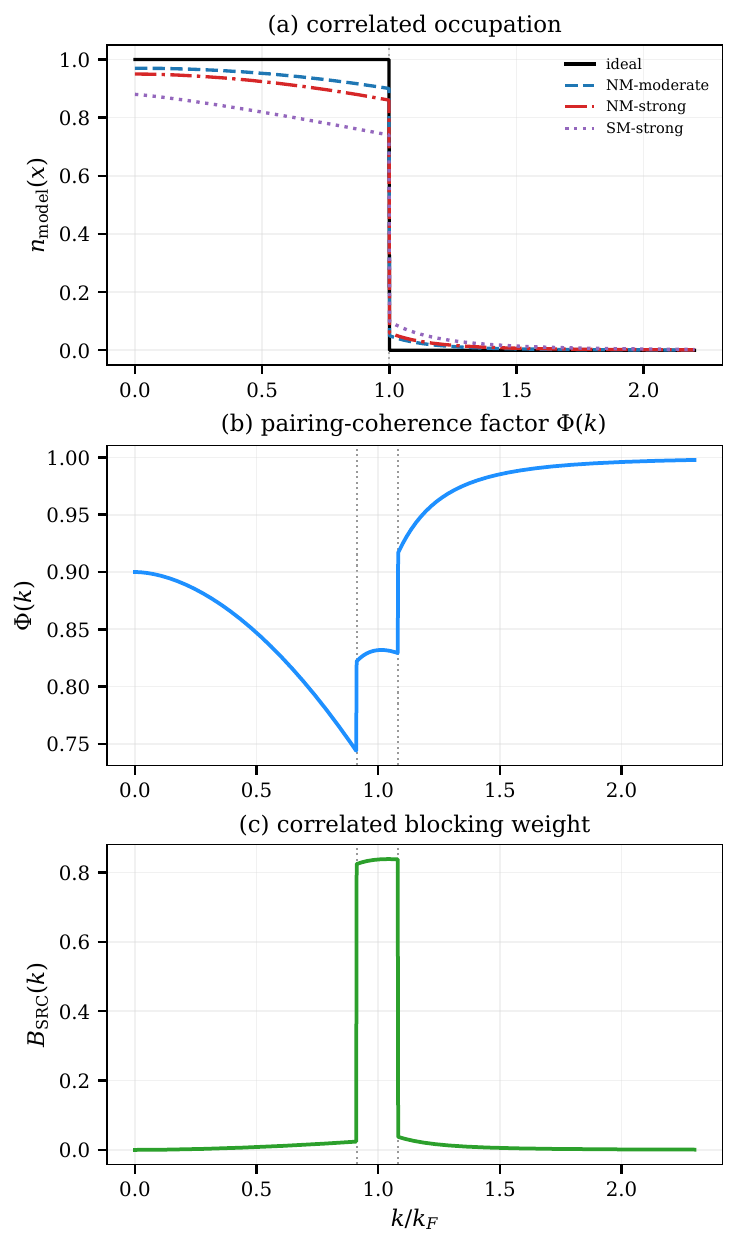}
    \caption{%
    Momentum-space inputs of the phenomenological construction.
    (a) Ideal and correlated occupation profiles
    $n_{\rm model}(x)$, with $x=k/k_{F\sigma}$, for the three benchmark
    parameter sets.
    (b) Pairing-coherence factor $\Phi(k)$ of Eq.~\eqref{eq:Phi}.
    (c) Correlated blocking weight $B_{\rm SRC}(k)$ of
    Eq.~\eqref{eq:Bsrc}.
    The vertical dotted lines in panels (b) and (c) indicate the two
    species-dependent Fermi momenta. The latter panels use the NM-strong
    occupation profile for the representative imbalanced configuration.
    }
    \label{fig:input-functions}
\end{figure}

\subsubsection{Correlated Pauli blocking: Model B}
\label{sec:modelB}

For ideal step occupations, the occupation mismatch itself is the
indicator of the interval between the two Fermi momenta,
\begin{equation}
B_0(k)=\Theta(k_{FB}-k)-\Theta(k_{FA}-k),
\end{equation}
equal to one for $k_{FA}<k<k_{FB}$ and zero otherwise.
With $k_{F\sigma}=\sqrt{2M\mu_\sigma}$
(Eq.~\eqref{eq:kfsigma-density}), the boundaries $k_{FA}$ and $k_{FB}$
bracket the spectral blocked interval $k_{1,2}=
\sqrt{2M(\mu\mp\sqrt{h^2-\Delta^2})}$, so that
$B_0(k)\,\Theta(h-E_k)=\Theta(h-E_k)$ holds identically for all
$\Delta\ge0$ whenever $h>\Delta$. The ideal model therefore reproduces
the blocking term of Refs.~\cite{Bedaque2003,Caldas2012} without an
additional momentum-window factor. Model B replaces
this hard interval with the occupation imbalance itself, evaluated with the
correlated distribution:
\begin{equation}
B_{\rm SRC}(k)=\bigl[n_B(k)-n_A(k)\bigr]_+,
\qquad
[z]_+\equiv\max(z,0).
\label{eq:Bsrc}
\end{equation}
Equation~\eqref{eq:Bsrc} is thus the direct occupation-based extension
of the ideal blocking indicator: in the sharp-step limit it is exactly
unity between $k_{FA}$ and $k_{FB}$ and zero outside. Its integrated
weight differs from that of $B_0(k)$. The reduced integrated weight
represents the overlap generated by depletion below the Fermi surfaces
and by their high-momentum tails. The spectral condition
for the blocked interval to exist, Eq.~\eqref{eq:blocked-condition}, and the
weight assigned to it are conceptually distinct. The former determines
where the lower quasiparticle branch is negative; the latter retains the
same occupation-mismatch interpretation after depletion and tail
formation. In the numerical implementation,
$B_{\rm SRC}(k)$ is evaluated using the exact positive part in
Eq.~\eqref{eq:Bsrc}.  
Separately, the Heaviside factor $\Theta(h-E_k)$ in
Eqs.~\eqref{eq:omega3d-ideal}--\eqref{eq:omega2d-ideal} is evaluated
through a smooth sigmoid gate for differentiability inside the numerical
solvers of Sec.~\ref{sec:numerics}.

The occupation input and the two momentum-dependent factors constructed
from it are illustrated in Fig.~\ref{fig:input-functions}. The occupation
profiles determine both the pairing-coherence factor $\Phi(k)$ and the
correlated blocking weight $B_{\rm SRC}(k)$, while the two quantities enter
different sectors of the grand potential.

\subsubsection{Four-model decomposition}
\label{sec:fourmodels}

The pairing and blocking sectors are switched independently, giving four
models compared throughout this work:
\begin{equation}
\begin{array}{c|cc}
\text{model} & \text{pairing sector} & \text{blocking sector}\\
\hline
\text{ideal} & \Phi=1 & B_0\\
P & \Phi\ \text{[Eq.~\eqref{eq:Phi}]} & B_0\\
B & 1 & B_{\rm SRC}\ \text{[Eq.~\eqref{eq:Bsrc}]}\\
P{+}B & \Phi\ \text{[Eq.~\eqref{eq:Phi}]} & B_{\rm SRC}\ \text{[Eq.~\eqref{eq:Bsrc}]}
\end{array}
\label{eq:fourmodels}
\end{equation}
Model P+B is the full realization of the phenomenological prescription,
while P and B are diagnostic switches that expose how the common occupation
input propagates through each sector separately. Both $\Phi(k)$ and $B_{\rm SRC}(k)$ recover the exact ideal limit.
The four-model comparison then determines the sign and magnitude of
each sector contribution and tests the stability of their ordering
across dimensions and occupation profiles.

\subsection{Three-dimensional formulation}
\label{sec:3dformulation}

Collecting Secs.~\ref{sec:ideal}--\ref{sec:fourmodels}, the 3D
correlated-occupation grand potential used by the calculation is
\begin{widetext}
\begin{equation}
\Omega_{3D}
=
\frac{\Delta^2}{g_{\rm eff}}
+
\int_0^{\Lambda}\frac{k^2dk}{2\pi^2}
\Bigl\{
\bigl[(\xi_k-\xi_k^{0})+\Phi(k)^2(\xi_k^{0}-E_k)\bigr]
-
B(k)\,(h-E_k)\,\Theta(h-E_k)
\Bigr\},
\label{eq:omega3d-full}
\end{equation}
\end{widetext}
with $\Phi(k)$ and $B(k)$ set according to Eq.~\eqref{eq:fourmodels}
(model ideal uses $\Phi=1$, $B=B_0$), $g_{\rm eff}$ fixed by
Eq.~\eqref{eq:geff} at reference gap $\Delta_0$, and $\Lambda=8$
(momentum units set by $M=1$) in the standard calculations.

Since $\Phi(k)$ and $B(k)$ are fixed functionals of the instantaneous
$(\mu,h)$-dependent occupations (Sec.~\ref{sec:src}) and do not depend on
$\Delta$,
the stationarity condition $\partial\Omega_{3D}/\partial\Delta=0$ can be
formed in closed form for all four models at once. Using
$\partial E_k/\partial\Delta=\Delta/E_k$ and dividing through by $\Delta$
(for $\Delta\neq0$),
\begin{equation}
\frac{2}{g_{\rm eff}}
=
\int_0^{\Lambda}\frac{k^2dk}{2\pi^2}
\left[
\frac{\Phi(k)^2}{E_k}
-
\frac{B(k)}{E_k}\,\Theta(h-E_k)
\right].
\label{eq:gapeq-3d}
\end{equation}
For the ideal model ($\Phi=1$, $B=B_0$, $h=0$) this reduces to
$2/g_{\rm eff}=\int_0^\Lambda k^2dk/2\pi^2\,/E_k$, i.e.\ exactly twice
Eq.~\eqref{eq:geff} evaluated at the solving $\Delta$, as required for
consistency with the definition of $g_{\rm eff}$ itself. For models P, B,
and P+B, Eq.~\eqref{eq:gapeq-3d} remains a well-defined closed equation for
$\Delta$, but $\Phi(k)$ and $B(k)$ are not simple enough functions of $k$
for it to be inverted algebraically; the paired branch is instead obtained
as the value of $\Delta$ minimizing $\Omega_{3D}(\Delta)$ on a dense
$\Delta$-grid with local refinement (Sec.~\ref{sec:numerics}), which
locates the same stationary point as Eq.~\eqref{eq:gapeq-3d} without
solving it symbolically. The normal branch is $\Delta=0$; the Sarma-like
branch, when present, is the local maximum of $\Omega_{3D}(\Delta)$
described in Sec.~\ref{sec:ideal}.

\subsection{Two-dimensional formulation}
\label{sec:2dformulation}

The corresponding 2D functional is
\begin{widetext}
\begin{equation}
\Omega_{2D}
=
\int_0^{\Lambda}\frac{k\,dk}{2\pi}
\left\{
\bigl[(\xi_k-\xi_k^{0})+\Phi(k)^2(\xi_k^{0}-E_k)\bigr]
+
\frac{\Delta^2}{2\epsilon_k+|\epsilon_B|}
\right\}
-
\int_0^{\Lambda}\frac{k\,dk}{2\pi}\,B(k)\,(h-E_k)\,\Theta(h-E_k),
\label{eq:omega2d-full}
\end{equation}
\end{widetext}
with $\Lambda=16$ in the standard calculations and $\epsilon_k=k^2/2M$ the free
dispersion of Eq.~\eqref{eq:omega2d-ideal}. A detail specific to two
dimensions is that the regularizing counterterm
$\Delta^2/(2\epsilon_k+|\epsilon_B|)$ is \emph{not} weighted by
$\Phi(k)^2$: it is bookkeeping for the ultraviolet divergence of the 2D
contact interaction, the direct analog of the momentum-independent term
$\Delta^2/g_{\rm eff}$ that sits outside the integral altogether in the 3D
functional \eqref{eq:omega3d-full}, and it is not a physical
occupation-weighted pairing contribution. Weighting it by $\Phi(k)^2$
together with the genuine anomalous term $\xi_k^0-E_k$ would suppress the
counterterm along with the pairing signal it is meant to regularize. Being
fixed by the vacuum two-body binding energy rather than by the
in-medium chemical potential, this counterterm is built from the free
dispersion $\epsilon_k$ and not from the shifted $\xi_k$ used everywhere
else in $\Omega_{2D}$; consequently $2\epsilon_k+|\epsilon_B|>0$ for all
$k\ge0$, and no finite-momentum pole can arise from this term.

The blocking structure mirrors the 3D case: the crossing condition
Eq.~\eqref{eq:blocked-condition} gives the blocked-interval boundary
$k_{1,2}^2=2M(\mu\pm\sqrt{h^2-\Delta^2})$ as in
Ref.~\cite{Caldas2012}, with the unit occupation mismatch of the ideal
theory replaced by $B_{\rm SRC}(k)$ of Eq.~\eqref{eq:Bsrc}.

By the same argument as in Sec.~\ref{sec:3dformulation} ($\Phi(k)$, $B(k)$
do not depend on $\Delta$), differentiating Eq.~\eqref{eq:omega2d-full} term
by term and dividing by $\Delta\neq0$ gives the closed stationarity
condition
\begin{equation}
0=
\int_0^{\Lambda}\frac{k\,dk}{2\pi}
\left[
-\frac{\Phi(k)^2}{E_k}
+
\frac{2}{2\epsilon_k+|\epsilon_B|}
+
\frac{B(k)}{E_k}\,\Theta(h-E_k)
\right]
\label{eq:gapeq-2d}
\end{equation}
for all four models. For the ideal model ($\Phi=1$, $B=B_0$),
Eq.~\eqref{eq:gapeq-2d} is identical to the gap equation of
Ref.~\cite{Caldas2012}, since $\Phi^2=\Phi=1$ there. For models P, B, and
P+B, $\Phi(k)^2$ and $B_{\rm SRC}(k)$ again prevent an algebraic inversion
for $\Delta$; the paired branch is located from the same $\Delta$-grid
minimization used in three dimensions (Sec.~\ref{sec:numerics}), which
satisfies Eq.~\eqref{eq:gapeq-2d} numerically rather than symbolically.
Equation~\eqref{eq:omega2d-full} is the functional used for the validation
described in Sec.~\ref{sec:results-phase}; no quantitative comparison
is presented in this section.

\subsection{Canonical thermodynamics}
\label{sec:canonical}

\subsubsection{Density reconstruction}
\label{sec:density-inversion}

Phase competition is a fixed-density problem, so densities must be
obtained from $\Omega$ rather than assigned by a free-gas formula. At fixed
$\Delta$,
\begin{equation}
n_T=n_A+n_B=-\frac{\partial\Omega}{\partial\mu},
\qquad
m=n_B-n_A=-\frac{\partial\Omega}{\partial h},
\label{eq:legendre-densities}
\end{equation}
Given target densities $n_A^\star,n_B^\star$ (equivalently
$n_T^\star=n_A^\star+n_B^\star$ and
$m^\star=n_B^\star-n_A^\star$), Eq.~\eqref{eq:legendre-densities} is
inverted to obtain the chemical potentials $(\mu,h)$ for each trial value of
$\Delta$ by minimizing the dimensionless residual
\begin{equation}
\mathbf{R}(\mu,h)=\left(
\frac{n_T-n_T^\star}{n_T^\star},
\frac{m-m^\star}{n_T^\star}
\right).
\label{eq:residual}
\end{equation} The inversion preserves the exact $A\leftrightarrow B$ symmetry and
does not use the ideal-gas relation between density and chemical potential as
a physical approximation. Finite-difference steps, solver tolerances, and
initialization procedures are specified in Sec.~\ref{sec:numerics}.

\subsubsection{Canonical energy}
\label{sec:canonical-energy}

For each fixed $(n_A,n_B,\Delta)$, having solved
Eq.~\eqref{eq:legendre-densities} for $(\mu,h)$, the canonical energy
density is the Legendre transform
\begin{equation}
E(n_T,m,\Delta)=\Omega(\mu,h,\Delta)+\mu n_T+h m.
\label{eq:legendre-energy}
\end{equation}
The normal branch is $E_N(n_A,n_B)\equiv E(n_T,m,0)$. The paired
(balanced) branch is
\begin{equation}
E_{\rm pair}(n)\equiv\min_{\Delta}E(2n,0,\Delta),
\end{equation}
obtained by minimizing the canonical energy over $\Delta$. The numerical
search and refinement strategy is described in Sec.~\ref{sec:numerics}. A Sarma-like branch, when it exists, is recorded as a local
\emph{maximum} of $\Omega(\Delta)$ at the corresponding fixed densities and
used only as a stability diagnostic, never as a candidate for the mixed
state below.

\subsubsection{Phase-separated state}
\label{sec:mixed}

The candidate mixed state consists of a normal fraction $x$ and a balanced
paired core of per-species density $n_c$, with the normal-shell densities
fixed by conservation of each species' total density:
\begin{equation}
n_A=(1-x)\,n_c+x\,n_A^{N},
\qquad
n_B=(1-x)\,n_c+x\,n_B^{N},
\label{eq:mixed-constraint}
\end{equation}
so that, for a given trial $(x,n_c)$,
\begin{equation}
n_A^{N}=\frac{n_A-(1-x)n_c}{x},
\qquad
n_B^{N}=\frac{n_B-(1-x)n_c}{x},
\label{eq:mixed-shell}
\end{equation}
subject to $n_A^N,n_B^N\ge0$. The mixed-state energy is
\begin{widetext}
\begin{equation}
E_{\rm mix}(n_A,n_B)
=
\min_{x,\,n_c}
\Bigl[
(1-x)\,E_{\rm pair}(n_c)
+
x\,E_N\bigl(n_A^{N}+n_B^{N},\,n_B^{N}-n_A^{N}\bigr)
\Bigr],
\label{eq:mixed-energy}
\end{equation}
\end{widetext}
using the same $E_{\rm pair}$ and $E_N$ of Sec.~\ref{sec:canonical-energy}.
The boundary cases $x=0$ (homogeneous balanced paired state) and $x=1$
(homogeneous normal state) are not counted as nontrivial phase separation;
a genuinely mixed ground state corresponds to $0<x<1$.

For numerical efficiency, the homogeneous paired and normal energies are
tabulated as functions of density before the minimization in
Eq.~\eqref{eq:mixed-energy}. Interpolation and minimization details are given
in Sec.~\ref{sec:numerics}.

\paragraph{Phase-boundary location.} At fixed coupling, the quantity
$\Delta E(p)\equiv E_{\rm mix}(p)-E_N(p)$, evaluated along a line of fixed
$n_T$ as a function of polarization $p=m/n_T$, is smooth and
non-increasing in the phase-separated region and identically zero once the
minimization in Eq.~\eqref{eq:mixed-energy} collapses to the trivial
$x=1$ branch. The critical polarization $p_c\equiv m_c/n_T$ is the boundary
between these two regimes. Numerically, $p_c$ is identified as the left edge of the interval in which
$E_{\rm mix}-E_N$ becomes zero; the bracketing and bisection procedure is
described in Sec.~\ref{sec:numerics}.

\section{Numerical implementation}
\label{sec:numerics}

The formalism is evaluated with numerical quadratures, nonlinear solvers,
interpolation, and convergence tests summarized below.

\paragraph{Momentum grid.}
The integrals in Eqs.~\eqref{eq:omega3d-full}--\eqref{eq:omega2d-full}
are evaluated by the trapezoidal rule. Because the correlated occupation
of Eq.~\eqref{eq:nmodel} is discontinuous at $q=1$, the integrands have
kinks at $k=k_{FA}$ and $k=k_{FB}$. The grid is therefore assembled from
uniform subintervals separated exactly at these momenta, preventing any
trapezoidal panel from crossing a discontinuity. The standard calculations
use $1600$ points in 3D with $\Lambda=8$ and $1800$ points in 2D with
$\Lambda=16$.

\paragraph{$\Delta$-grid.}
The paired branch of Sec.~\ref{sec:canonical-energy} is located on a
$120$-point $\Delta$ grid, followed by bounded scalar refinement around
each robust local minimum.

\paragraph{Density inversion.}
Equation~\eqref{eq:residual} is solved with scaled-residual tolerances of
$3\times10^{-3}$ in 2D and $5\times10^{-3}$ in 3D. If the initial attempts
fail, a $9\times9$ (2D) or $7\times7$ (3D) seed grid in $(\mu,h)$ is
scored, the best candidates are refined by bounded least squares, and a
differential-evolution search is used only if necessary.

\paragraph{Mixed-phase minimization.}
The search in $(x,n_c)$ uses a $101\times101$ grid followed by a
Nelder--Mead refinement. The critical polarization is located by bisection
to $10^{-5}$--$10^{-6}$ in $p$. This tolerance controls only the edge of
the zero plateau in $E_{\rm mix}-E_N$; the total uncertainty in $p_c$ is
set mainly by density inversion and interpolation.

\paragraph{Failure handling.}
Density-table points that fail the tolerance of
Sec.~\ref{sec:density-inversion} are excluded rather than extrapolated.
Such failures occur only at extreme density ratios, outside the region used
to determine the phase boundary.

\paragraph{Consistency checks.}
The implementation preserves the $\Delta=0$ independence of the normal
energy from $\Phi(k)$ and the invariance of the canonical energy under
simultaneous exchange of the species labels and $h\to-h$.

\paragraph{Thermodynamic derivatives.}
The densities are obtained from centered finite differences of
$\Omega(\mu,h,\Delta)$ at fixed $\Delta$. Since
$k_{F\sigma}=\sqrt{2M\mu_\sigma}$ depends on $(\mu,h)$, the variation of
$n_\sigma(k)$ is included in the total derivative. Halving or doubling the
finite-difference step within
$[5\times10^{-4},2\times10^{-3}]$ leaves the reconstructed densities
unchanged within numerical accuracy.

\paragraph{Cutoff convergence.}
In 3D, increasing the cutoff from $\Lambda=8$ to $\Lambda=20$ changes the
balanced gap and the canonical energies by less than $0.5\%$ at the
strongest sampled coupling. At the weaker representative point, the
P+B-to-ideal gap ratio changes by approximately $1.9\%$, still within the
overall numerical uncertainty. In 2D, where the integral is ultraviolet
finite after renormalization through $\epsilon_B$, extending the numerical
limit beyond $\Lambda=16$ changes $p_c$ by less than
$3\times10^{-3}$. The visually uneven spacing of the curves in
Fig.~\ref{fig:phase-diagrams} reflects piecewise-linear rendering of a
finite coupling grid; the computed sequences remain monotonic.

\paragraph{Resolution refinement for $p_c$.}
At one representative coupling in each dimension, the calculation was
repeated with finer momentum, mixed-phase, and polarization grids and
tighter density-inversion tolerances. The ideal--B separation remains
resolved in both dimensions. The P--(P+B) separation remains finite in 2D
but vanishes within numerical resolution in 3D, where the two models are
indistinguishable.

\paragraph{Parameter-set sensitivity.}
The alternative profiles in Table~\ref{tab:benchmarks} preserve the
ordering ideal$\,\approx\,$B and P$\,\approx\,$P+B in both dimensions.
The displacement from the ideal boundary grows with the strength of the
prescribed occupation modification. The reported phase boundaries use the
\textit{NM-strong} profile as the common 2D--3D reference.

\section{Results}
\label{sec:results}

\subsection{Grand-potential landscape and homogeneous branches}
\label{sec:results-omega}

\begin{figure*}
\centering
\includegraphics[width=\textwidth]{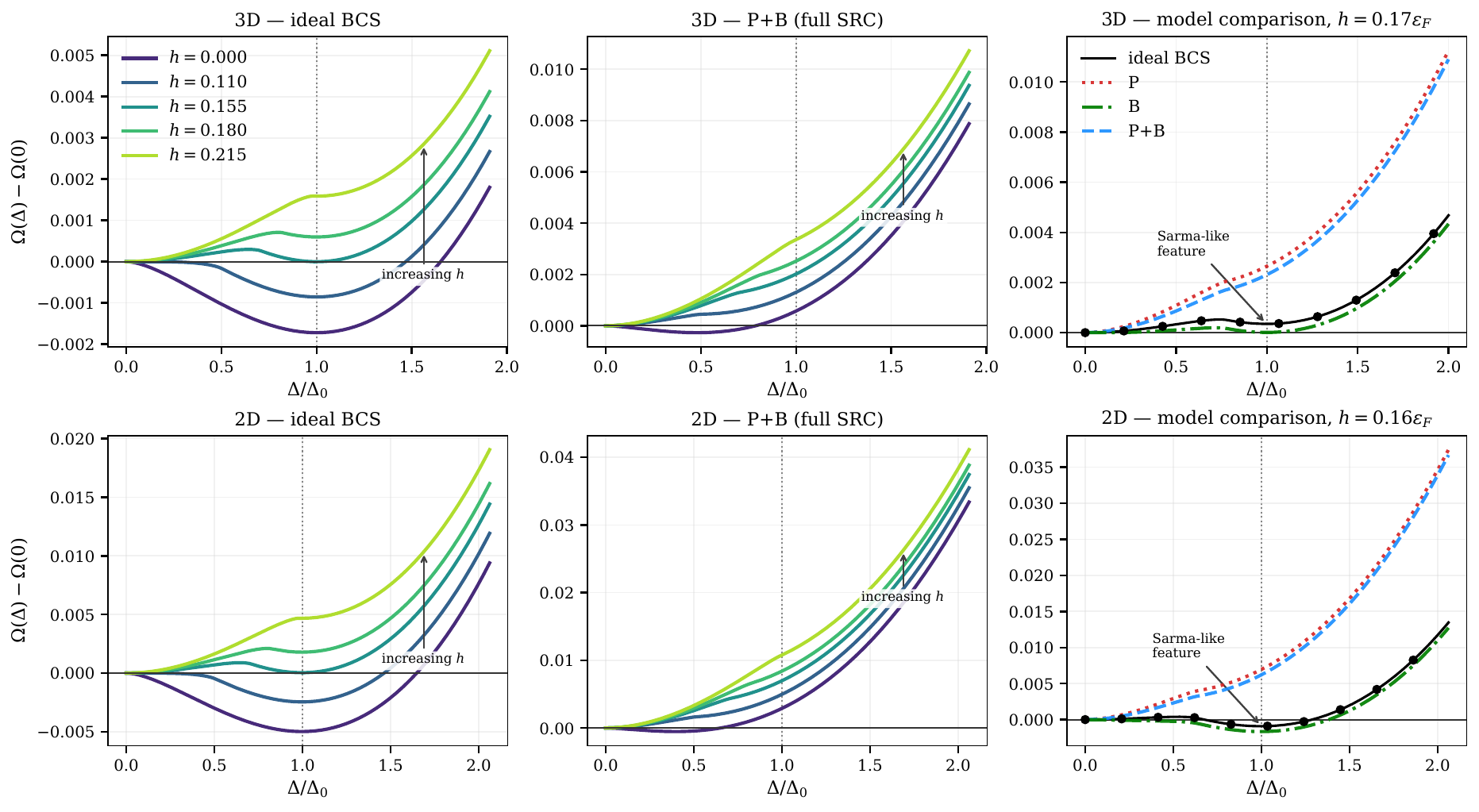}
\caption{Grand-potential landscape for the ideal and SRC-modified models in
three (upper row) and two (lower row) dimensions, shown as
$\Omega(\Delta)-\Omega(0)$ with $\Delta$ normalized by the reference gap
$\Delta_0$ ($\Delta_0=0.22$ in 3D, $\Delta_0\approx0.252$ in 2D). The
first column shows the ideal model for five values of the imbalance field
$h$ (legend); the second shows the P+B model at the same $h/\Delta_0$
ratios; the third compares the ideal, P, B, and P+B models at one fixed
$h$ (3D: $h=0.17$; 2D: $h=0.16$). The vertical dotted line marks
$\Delta/\Delta_0=1$; the horizontal line marks the normal-state reference,
$\Omega-\Omega(0)=0$. The arrow labeled ``Sarma-like feature'' marks a
homogeneous, fixed-$(\mu,h)$ stationary point of $\Omega(\Delta)$, distinct
from the fixed-density canonical ground state of Sec.~\ref{sec:mixed}.}
\label{fig:omega-grid}
\end{figure*}

Figure~\ref{fig:omega-grid} shows
$\Omega(\Delta)-\Omega(0)$ for the ideal and SRC-modified models.
The fixed-$(\mu,h)$ reference values were specified in
Sec.~\ref{sec:framework}. 
These choices set the plotting scale and do not enter the canonical
phase-boundary calculations. The ideal model (first column) sets the
reference structure: at $h=0$ the potential has a single minimum at
$\Delta=\Delta_0$, by construction of the reference coupling
(Sec.~\ref{sec:ideal}). As $h$ increases, this minimum is progressively
raised relative to the normal state and eventually becomes only locally
stable, developing a local maximum followed by a local minimum before the
potential turns up further -- a three-extremum, homogeneous fixed-$(\mu,h)$
signature of the Sarma-type competition discussed in Sec.~\ref{sec:ideal}.
Consistent with that discussion, this intermediate stationary point is not
a candidate fixed-density ground state; the actual ground state at fixed
density follows instead from the canonical mixed-state construction of
Sec.~\ref{sec:mixed} and is shown in Fig.~\ref{fig:phase-diagrams}.

The third-column panels compare the four models at fixed $h$. Activating
the pairing-coherence correction (P) raises the finite-$\Delta$ branch
well above the ideal curve -- in 2D enough to turn a locally favorable
paired branch into a locally unfavorable one. The blocking-only model (B) stays close to the ideal curve, whereas the
combined model (P+B) tracks P closely. In 2D the small offsets between the
members of each pair are numerically resolved; in 3D they are comparable to,
or below, the standard numerical resolution. The robust result is therefore the
pairwise ordering $\text{ideal}\approx B$ and $P\approx\text{P+B}$, with
the displacement between these two pairs controlled by P in both
dimensionalities.

\subsection{Critical polarization and phase boundaries}
\label{sec:results-phase}

\begin{figure*}
\centering
\includegraphics[width=\textwidth]{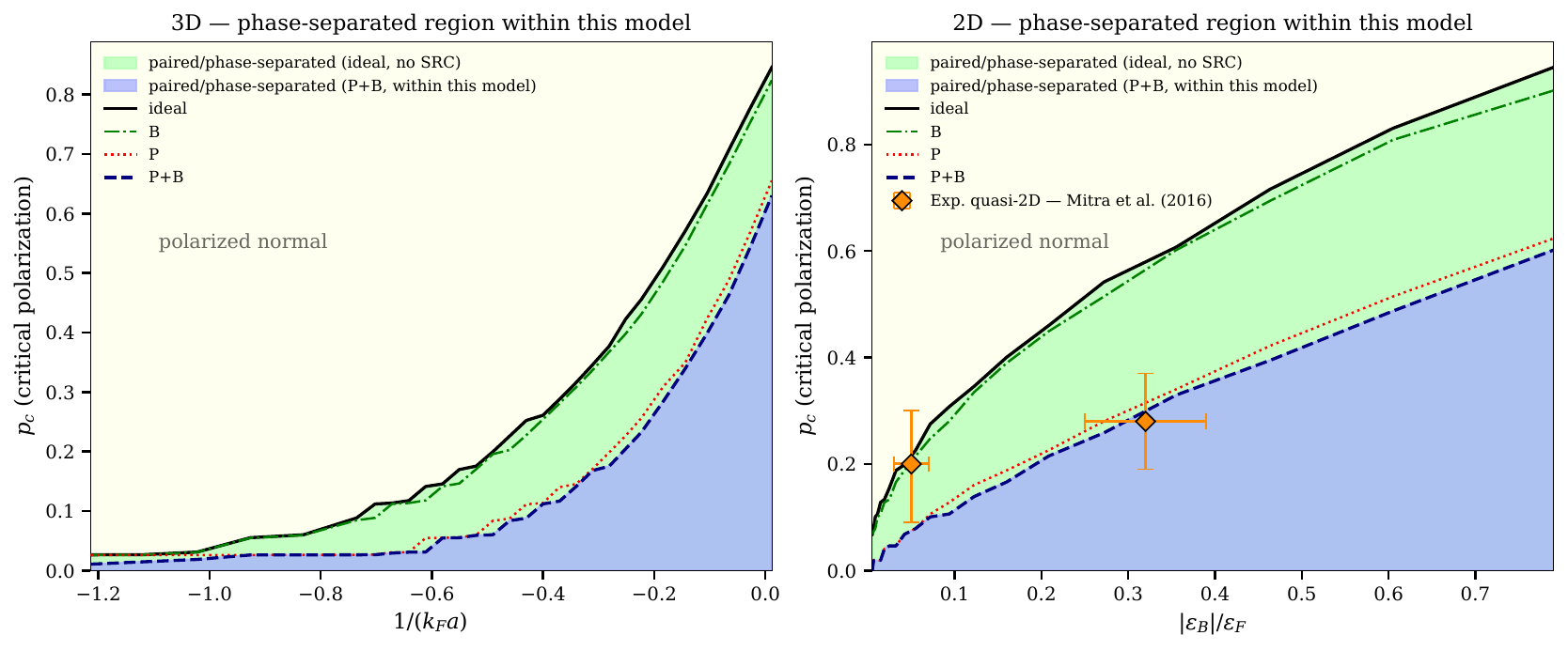}
\caption{Critical polarization $p_c\equiv m_c/n_T$
(Sec.~\ref{sec:mixed}) separating the paired/phase-separated region
($p<p_c$) from the polarized normal state ($p>p_c$), obtained from the
canonical mixed-state construction of Sec.~\ref{sec:mixed}. Left: 3D
result as a function of $1/(k_Fa)$ (Eq.~\eqref{eq:invkfa}), the standard
BCS--BEC crossover coupling. Right: 2D result as a function of
$|\epsilon_B|/\epsilon_F$. In both panels, the coupling strength increases
from left to right. Curves show the ideal, P, B, and P+B models; the
light-green region is bounded by the ideal curve, while the dark-blue
region is bounded by P+B and marks the paired/phase-separated domain that
remains within the full prescription. Filled diamonds show the global
critical polarizations reported by Mitra et~al.~\cite{Mitra2016} for the
disappearance of the balanced central core in a trapped quasi-2D
$^6$Li gas at $E_B/E_F=0.05(2)$ and $0.32(7)$. The experimental notation
$E_B$ and $E_F$ is retained from Ref.~\cite{Mitra2016}. The points are
placed at their reported numerical values of $E_B/E_F$ only to provide a
reference interaction scale; because the experimental and homogeneous
Fermi energies are defined differently, their horizontal positions should
not be interpreted as a direct identification with
$|\epsilon_B|/\epsilon_F$. Error bars denote the experimental
uncertainties. The measurements were performed at
$T/T_F=k_BT/\mu_{0\uparrow}=0.18(5)$ using the experimental
majority-component definition of $T_F$. The data are therefore shown only
as a semiquantitative reference for the polarization scale and its
interaction dependence, not as measurements of the same homogeneous
$T=0$ canonical boundary represented by the curves.}
\label{fig:phase-diagrams}
\end{figure*}

Figure~\ref{fig:phase-diagrams} shows the critical polarization $p_c\equiv
m_c/n_T$ from the same canonical construction in both dimensionalities,
plotted directly in these dimensionless units at fixed $n_T$ per scan.
Below the plotted curve the mixed-state minimization of
Eq.~(\ref{eq:mixed-energy}) favors the paired/phase-separated state; above
it, the polarized normal state is favored (within the restricted
phase space of uniform paired, phase-separated, and normal
configurations; spatially modulated states are not included).

\paragraph{3D result.}
After applying the cutoff-consistent conversion of
Eq.~\eqref{eq:invkfa}, the sampled range is
$1/(k_Fa)\approx-1.20$ to $0.00$, extending from the BCS regime toward
unitarity. The ideal and B boundaries remain numerically close throughout
this range, while the P and P+B boundaries form a second, substantially
lower pair. At $1/(k_Fa)\simeq-0.74$, for example, the standard
calculation gives $p_c=0.0883$ (ideal), $0.0845$ (B), $0.0265$ (P), and
$0.0264$ (P+B). The ideal--B separation is small
($\approx4\%$ of the ideal value) but remains resolved under the numerical
refinement described in Sec.~\ref{sec:numerics}; P and P+B are numerically
indistinguishable under the same refinement. The 3D result therefore
consists of a small but resolved ideal--B separation and a much larger
displacement between the ideal/B and P/P+B pairs.

\paragraph{2D result.}
The ideal boundary reproduces the qualitative trend of the published
two-dimensional result of Ref.~\cite{Caldas2012}, namely a monotonic
increase of $p_c$ with $|\epsilon_B|/\epsilon_F$ and a comparable
curvature; no quantitative overlay is attempted. The same hierarchy as
in 3D is observed: ideal and B remain close, whereas P and P+B form a
second, substantially lower pair. At
$|\epsilon_B|/\epsilon_F\approx0.27$, the 2D-normalized calculation gives
$p_c=0.540$ (ideal), $0.514$ (B), $0.281$ (P), and $0.259$ (P+B).
The pairing-sector correction accounts for approximately $92\%$ of the
total reduction relative to the ideal boundary, while the blocking-only
shift corresponds to approximately $4.7\%$ of the ideal value. The P and
P+B boundaries differ by approximately $0.022$, a larger separation than
in 3D, but remain much closer to each other than to the ideal and B
curves.

For reference, Fig.~\ref{fig:phase-diagrams} also includes the critical
polarizations reported by Mitra et~al.~\cite{Mitra2016} for a trapped
quasi-two-dimensional $^6$Li gas. We retain the experimental notation
$E_B$ and $E_F$ used in that work, whereas $\epsilon_B$ and $\epsilon_F$
denote the binding and Fermi-energy scales of the present homogeneous
calculation. The experimental points are therefore shown only as a
semiquantitative reference. They correspond to the global trap
polarization at which the balanced central core disappears, measured at
finite temperature, $T/T_F=0.18(5)$, while our curves represent the
$T=0$ canonical boundary of a homogeneous system. In addition, the
experiment is quasi-2D and uses a trap-based definition of $E_F$, so the
comparison concerns the characteristic polarization scale and its
interaction dependence rather than an equality of the underlying
dimensionless coupling variables.

At $E_B/E_F=0.05(2)$, the experimental point lies closest to the ideal
and B curves, whereas at $E_B/E_F=0.32(7)$ the P and P+B curves lie
closer to the measured value. With only two experimental points in the
overlapping interaction range, no model is uniquely selected. The data
provide a qualitative constraint on the interaction dependence of $p_c$
but do not constitute a quantitative validation of any of the four
prescriptions.

\paragraph{Dimensional comparison.}
Both dimensionalities exhibit the same hierarchy,
$\text{ideal}\approx B$ and $P\approx\text{P+B}$, with the pairing-sector
modification producing the dominant displacement of $p_c$ and the blocking
correction remaining small, at about $4\%$ of the ideal value in 3D and
$4.7\%$ in 2D. The balanced gaps show the same pairing, since B acts only
in the imbalance sector of $\Omega$ (Sec.~\ref{sec:modelB}). Because the
two panels use different coupling variables, the comparison concerns this
common hierarchy rather than a point-by-point correspondence. Their
different curvatures reflect the underlying interaction scales: in 3D,
$\Delta/\epsilon_F\propto\exp[\pi/(2k_Fa)]$ in the weak-coupling BCS regime,
leading to a slow initial rise followed by a sharper increase, whereas in
2D the existence of a bound state at arbitrarily weak attraction and
$\Delta^2\propto\epsilon_F|\epsilon_B|$ produce a faster initial increase
and a progressively weaker growth.

\section{Discussion}
\label{sec:discussion}

The four-model construction yields the recurring hierarchy
\begin{equation}
\text{ideal}\simeq B,
\qquad
P\simeq P{+}B,
\end{equation}
in both the grand-potential landscapes of
Fig.~\ref{fig:omega-grid} and the canonical phase boundaries of
Fig.~\ref{fig:phase-diagrams}. Most of the P+B displacement is already
generated by P, while B adds a smaller correction in the same direction.
This ordering persists in both dimensionalities and for all occupation
profiles tested. At the representative 2D coupling, adding B to P changes
$p_c$ by approximately $0.022$; in 3D, P and P+B become numerically
indistinguishable under refinement.

The hierarchy is consistent with the momentum domains sampled by the two
prescriptions. The pairing weight $\Phi(k)^2$ acts throughout the anomalous
contribution and remains active at $h=0$, whereas $B_{\rm SRC}(k)$ acts only
within the blocked interval that opens for $h>\Delta$. This distinction
explains why a larger response from P is plausible, although the sign and
magnitude of both corrections remain numerical results of the model.

Microscopic pairing calculations provide the relevant physical context.
Correlated-basis-function studies show substantial changes in the
$^{1}S_{0}$ gap relative to uncorrelated BCS theory
\cite{Fabrocini2008}, while self-consistent spectral-function approaches
find that SRC fragment single-particle strength and reduce the coherent
weight available for pairing \cite{Ding2016}. Rios, Polls, and Dickhoff
similarly obtained suppressed pairing gaps in several nuclear channels
when SRC-modified spectral strength was included \cite{Rios2017}. These
results associate the pairing response not only with the integrated
high-momentum population, but also with the depletion and fragmentation of
strength in the momentum-energy region sampled by the pairing kernel.

More generally, the gap depends on the quasiparticle spectrum,
single-particle residue, effective interaction, and medium polarization
\cite{Sedrakian2019}. Correlations can also modify the zero-temperature
gap, the critical temperature, and their relation \cite{Drissi2022}.
Their effect therefore cannot be represented in general by a universal
rescaling of the BCS gap.

Model P reproduces only the qualitative direction of this microscopic
response. It reduces the occupation-dependent weight of the anomalous
sector, but does not include the correlated spectral dynamics of
Refs.~\cite{Fabrocini2008,Ding2016,Rios2017,Sedrakian2019,Drissi2022}:
$n_\sigma(k)$ is prescribed, the quasiparticle spectrum remains
mean-field-like, and $\Phi(k)$ is a phenomenological weight rather than a
kernel derived from dressed propagators. Within this restricted
description, weakening the anomalous sector lowers the polarization range
over which paired or phase-separated states remain favorable.

The cited microscopic studies also do not imply $P\gg B$, since they do
not separate anomalous and imbalance-blocking responses. That decomposition
is specific to the present framework. Model P leaves the $\Delta=0$ normal
reference unchanged and isolates the anomalous condensation contribution,
whereas model B replaces the ideal unit-weight blocked interval by the
correlated occupation mismatch. The four-model comparison therefore
separates two responses that would otherwise be absorbed into a single
effective correction.

Pairing and high-momentum structure may also influence each other.
Li \textit{et al.} showed that neutron--proton pairing can modify the
high-momentum distribution \cite{Li2026}. Such feedback is absent here
because $n_\sigma(k)$ is fixed before minimization. A self-consistent
extension would determine the spectral strength, momentum distribution,
and pairing field together. Within the present sector-resolved language,
such a treatment would let $n_\sigma(k)$ respond to $\Delta$ and $h$
rather than being prescribed in advance, so that $\Phi(k)$ and
$B_{\rm SRC}(k)$ would themselves depend on the solution instead of on an
input profile. This would test whether the dominance of the pairing
sector found here persists once the occupation is allowed to react to
the pairing field, or whether it is partly an artifact of holding
$n_\sigma(k)$ fixed, and it would let the depletion and high-momentum
tail narrow or broaden as $h$ approaches the pairing-breaking threshold,
a response the present construction cannot capture since $n_\sigma(k)$
carries no memory of $\Delta$.

A related extension concerns pairing states with finite center-of-mass
momentum. The present framework retains only the uniform BCS order
parameter and the bulk phase-separated mixture of Sec.~\ref{sec:mixed};
it does not include Fulde--Ferrell--Larkin--Ovchinnikov (FFLO) pairing or
the associated fluctuations. Beyond-mean-field treatments of polarized
Fermi gases find that FFLO correlations and the associated quantum
criticality reshape the phase diagram and the quasiparticle spectrum in
ways not captured by the uniform paired/normal/mixed competition
considered here \cite{Pini2023,Pirolo2026}. Comparing the present
SRC-inspired sector decomposition with an FFLO-including treatment,
quantitatively through the resulting phase boundary and qualitatively
through the critical behavior near the normal-to-paired transition, is a
natural direction left for future work.

The main limitations follow from the phenomenological and mean-field
construction. The calculation is restricted to $T=0$, and the implicit
$(\mu,h)$ dependence of
$k_{F\sigma}=\sqrt{2M\mu_\sigma}$ makes the density response part of the
prescription rather than the derivative of a conserving microscopic
functional. Finite temperature would smooth the occupation discontinuities
and the blocked interval, while pairing fluctuations and spatially
modulated phases could alter the phase-separated region, particularly in
2D. The mixed-state construction of Sec.~\ref{sec:mixed} also treats the
paired and normal domains as bulk phases joined at a sharp interface of
negligible energy; a finite surface tension between the superfluid and
normal regions is a long-standing and still only partially resolved
ingredient of the phase diagram of strongly interacting,
population-imbalanced Fermi gases
\cite{DeSilvaMueller2006,Caldas2007,BaurEtAl2009}, and its inclusion could
shift the reconstructed phase boundary away from the simple mixing rule
adopted here. Figure~\ref{fig:phase-diagrams} should therefore be
interpreted as a $T=0$, uniform-pairing baseline within the stated model.

\section{Conclusions}
\label{sec:conclusions}

We introduced a sector-resolved phenomenological framework in which the
same SRC-inspired occupation modifies the anomalous pairing contribution,
the imbalance-blocking term, or both. The resulting ideal, P, B, and P+B
models were compared through grand-potential landscapes and a full
fixed-density reconstruction of the phase boundary in two and three
dimensions.

The canonical results consistently satisfy
$\text{ideal}\simeq B$ and $P\simeq P{+}B$. The pairing-sector
modification produces most of the reduction in critical polarization,
whereas the blocking contribution remains a smaller same-sign correction,
amounting to approximately $4\%$ of the ideal value in 3D and $4.7\%$ in
2D. This ordering persists across both dimensionalities and the occupation
profiles tested.

The framework therefore provides a compact diagnostic of how a prescribed
correlated occupation propagates through distinct sectors of an imbalanced
superfluid. It also defines a baseline for extensions including finite
temperature, spatially modulated phases, and self-consistent occupations
and quasiparticle spectra.

\bibliographystyle{apsrev4-2}
\bibliography{references}

\end{document}